# Quantum Spin Liquids

C. Broholm[1], R. J. Cava[2], S. A. Kivelson[3], D. G. Nocera[4], M. R. Norman[5], T. Senthil[6]

[1]Institute for Quantum Matter and Department of Physics and Astronomy, The John Hopkins University, Baltimore, MD 21218
[2]Department of Chemistry, Princeton University, Princeton, NJ 08544
[3]Department of Physics, Stanford University, Stanford, CA 94305
[4]Department of Chemistry and Chemical Biology, Harvard University, Cambridge, MA 02138
[5]Materials Science Division, Argonne National Laboratory, Argonne, IL 60439
[6]Department of Physics, Massachusetts Institute of Technology, Cambridge, MA 02139

**Spin liquids are quantum phases of matter that exhibit a variety of novel features associated with their topological character. These include various forms of fractionalization – elementary excitations that behave as fractions of an electron. While there is not yet entirely convincing experimental evidence that any particular material has a spin liquid ground state, in the past few years, increasing evidence has accumulated for a number of materials suggesting that they have characteristics strongly reminiscent of those expected for a quantum spin liquid.**

In the early days of quantum mechanics, Heisenberg and others achieved an understanding of ferromagnetism by considering a state in which all the spins point in a single direction. Such a state is an eigenstate of the spin operator $S_z$. This is in contrast to Louis Néel's proposal of a lattice of alternating spins (an antiferromagnet) which as a consequence promoted great controversy at the time of its introduction (*1*). But it is now understood that the antiferromagnetic ground state is a classic example of the ubiquitous phenomenon of spontaneously broken global symmetry: the ground state has lower symmetry than the underlying Hamiltonian. The broken symmetry point of view enables understanding of a number of universal properties of the antiferromagnetic state, and shows their unity with similar phenomena in other ordered phases of matter. The same ideas when imported into high energy physics underlie many of the successes of the Standard Model. For phases of magnetic matter, it is now known that a variety of different kinds of spatially oscillating magnetic ordering patterns are possible, each corresponding to distinct broken symmetries. Still, the idea of a "quantum spin liquid", in which local moments do not freeze and there is no symmetry breaking (in the Landau sense), has long been discussed (*2*). Such states were contemplated in 1973 when Philip Anderson proposed that the ground state of the antiferromagnetic near-neighbor Heisenberg model on a triangular lattice might be a spin liquid. Specifically, he introduced the resonating valence bond picture of a spin liquid wavefunction, based on the resonating single/double C-C bonds developed by Linus Pauling to explain the electronic structure of benzene rings (*3*). Anderson's paper languished in relative obscurity until he resurrected the idea in the context of the newly discovered high temperature cuprate superconductors at the beginning of 1987 (*4*). It was realized soon afterwards by Kivelson, Rokhsar and Sethna (*5*) that the excitations of the spin liquid are topological in nature, and by Kalmeyer and Laughlin (*6*) that a version of the spin liquid could be constructed as a spin analogue of the celebrated fractional quantum Hall state.

These developments in 1987 led to an explosion of interest in quantum spin liquids that continues to this day.  In common with the fractional quantum Hall states, but distinct from conventional ordered states, the theory of the quantum spin liquid brings new concepts such as emergent gauge fields into condensed matter physics.  It is not our intent here to cover this work in any great depth, as there exist several excellent reviews (*7,8,9,10,11*).  Rather, we wish to take a broader look at the field.  In particular, what are the big questions, both in theory and experiment?

*What are quantum spin liquids?*  To discuss them in the clearest context, let us focus on the idealized situation of quantum spins arranged in a periodic crystalline lattice, with interactions that are short ranged in space.  This set-up describes correctly the essential physics of Mott (i.e. interaction driven) insulating materials.  In much of the literature, Mott insulating materials that do not magnetically order down to temperatures at which the spin dynamics is clearly quantum mechanical (i.e. much below the measured Curie-Weiss temperature) are considered to be spin liquid candidates.  Yet such a definition is not very enlightening, as it includes ground states that are not spin liquids, including conventional quantum paramagnets (e.g. with an even integer number of electrons per unit cell) and valence bond solids (*12*) (i.e., a lattice of spin singlets).  The latter in principle can be detected through the associated translation-symmetry-breaking, which can, however, be arbitrarily weak and masked by quenched disorder.  A more precise characterization comes from considering the structure of many particle quantum entanglement in the ground state. The prototypical ground state wavefunction of conventional magnetically ordered states of matter, including those mentioned above, is simply described by specifying the state of each spin in the lattice. The ability to independently specify the quantum state of individual parts of a quantum many particle system requires that the different parts have no essential quantum entanglement with each other.  Thus, the prototypical ground state wavefunctions for conventional states of magnetic matter may be said to have short range quantum entanglement between local degrees of freedom.  In contrast, the quantum spin liquid refers to ground states in which the prototypical wavefunction has long range quantum entanglement between local degrees of freedom (Figure 1d).  Under smooth deformations, such a wavefunction cannot be reduced to a product state wavefunction in real space (more technically, the ground state wavefunction cannot be completely disentangled into a product state with a finite depth circuit made up of local unitary transformations).  Such long-range quantum entanglement should be distinguished from the more familiar long-range order that characterizes broken symmetry phases.  Thus, the quantum spin liquid is a qualitatively new kind of ground state compared to conventional short range entangled states.

Just as there is no single type of magnetic order, there is no single type of quantum spin liquid.  Loosely speaking, they correspond to different patterns of long-range entanglement.  In addition, a useful (but coarse) classification distinguishes two classes of spin liquids based on whether the excitation spectrum above the ground state has a gap or not.  Gapped spin liquids are simpler theoretically, and are characterized by the global topological structure of their ground state wavefunctions.  Thus, they are said to have 'topological order', a concept that also describes fractional quantum Hall systems.  Such gapped spin liquids have well-defined emergent quasiparticles.  However, a remarkable feature is that these quasiparticles carry a topological signature that prevents them from being created in isolation. They can only be created in non-topological multiplets, which can be pulled apart to yield multiple individual quasiparticles. A single isolated quasiparticle thus represents a non-local disturbance of the ground state. This non-locality means it can be detected far away by operations that involve moving other emergent quasiparticles. Thus, quasiparticle excitations have non-local "statistical" interactions (like a charge moving around a magnetic flux). In two space dimensions, this implies the quasiparticles are anyons with non-trivial statistics when they braid around each other (by an anyon, we mean any quasiparticle that has non-trivial statistical interactions either with itself or with other quasiparticles; it is certainly possible for an anyon to have bosonic self-statistics, and yet non-trivial mutual statistics with some other quasiparticle).

There is a rich, formal theory of anyons in such topological ordered phases. In three space dimensions, apart from emergent point-like quasiparticles, there are also loop-like excitations with a line tension. There is a non-trivial braiding phase when a quasiparticle encircles a loop excitation. In either case, the non-locality associated with the quasiparticle excitation enables it to carry fractional quantum numbers associated with a global symmetry. A typical example of such a quasiparticle – known as a spinon – carries a spin ½ and charge 0 (Figure 1a). In contrast, local excitations in any insulating magnet must necessarily carry integer spin.

A second distinct class of spin liquids have a gapless excitation spectrum. In the simplest example of such phases, the gapless spectrum admits a quasiparticle description. However, interestingly, this is not guaranteed. There also are gapless spin liquid phases where the quasiparticle description completely breaks down (*13*). In general, gapless spin liquids have power law correlations of physical measurable quantities.

Given this variety of quantum spin liquid phases, what is the best theoretical framework in which we should think about them? Over the years, it has become clear that a powerful and convenient framework is provided by low energy effective theories that involve emergent gauge fields (*14,15,16,17*). Specifically, the low energy effective theory of a quantum spin liquid is a *deconfined* gauge theory. The gauge theory description elegantly captures the non-local entanglement and its consequences.

It is instructive to consider a prominent example of a quantum spin liquid phase described by an emergent deconfined Z2 gauge field (*18,19,20,21*). In two space dimensions, the excitations consist of a gapped excitation (the e particle) that carries Z2 gauge charge and another gapped excitation (the m particle) that carries Z2 gauge flux. These two have a long ranged statistical interaction: the wavefunction changes sign when an e particle is taken around an m particle. It is also possible to have a bound state of e and m (denoted ε). The e and m have bosonic statistics, however their mutual braiding phase implies that ε has fermionic statistics (Figure 1c). When such a phase occurs in lattices of an odd number of spin-½ local moments with spin rotation symmetry, it can be shown that the e particle carries spin-½ (and is a spinon with bosonic statistics). Further, the m particle has spin 0 but will carry fractional crystal momentum. It is known as the vison (Figure 1b). As their bound state, the ε particle, also carries spin ½, it is known as the fermionic spinon.

There are multiple ways of thinking about how a phase with such an excitation structure might come about. A close and very useful analogy is with the excitations of the familiar BCS s-wave superconductor. The excitations of a superconductor include the Bogoliubov quasiparticle and quantized vortices associated with h/2e magnetic flux. It is convenient to think about it in a basis where it is formally electrically neutral. Then it has a braiding phase π with the h/2e vortex. The Z2 quantum spin liquid may be viewed as a phase-disordered version of a superconductor where long range order is destroyed by quantum phase fluctuations. In this description, the fermionic spinon is identified as the cousin of the Bogoliubov quasiparticle (*18,22,23*), while the vison is identified as the cousin of the h/2e vortex (*18,23*). The close relationship between the Z2 spin liquid and the superconductor suggests that, if a spin liquid Mott insulator is found in a material, then doping it might naturally lead to superconductivity. Indeed, this is the original dream of the RVB theory as a mechanism for high temperature superconductivity.

Other quantum spin liquid phases will have other emergent gauge groups (for example, the familiar U(1) gauge field; these are not obviously connected to superconductivity in any simple way). Given the importance of the gauge theory description, it is not surprising that many concepts in high energy physics have been realized in the spin liquid context, including monopole-like excitations, which have been proposed in the context of the 3D pyrochlore lattice (*24*). Furthermore, it is conceptually straightforward to combine features of a spin liquid with more conventional phases, giving rise to additional new quantum phases of matter with combined topological order and broken symmetries (*25,26*), or even novel metallic phases with a Fermi surface whose enclosed volume violates Luttinger's theorem (that is, it is not proportional to the electronic density) (*27*).

*Do quantum spin liquids exist in theory?* This question was settled in a variety of different ways in the late 1980s and 1990s, when the first stable effective field theory descriptions of both the Z2 quantum spin liquid (*18,19,20,21*), and a different time reversal broken version (known as a chiral spin liquid) (*28*), were developed and their physical properties elucidated. Specific models that realize the Z2 spin liquid were constructed in a large-N generalization of the SU(2) Heisenberg magnet on square lattices with short ranged interactions involving more than just nearest neighbors (*20*) (so as to frustrate classical order), and on frustrated non-bipartite lattices (e.g. the triangular and kagome lattices (*29*)). A Z2 topological ordered state was also shown to be present in the quantum dimer model on the triangular lattice (*30*). And Kitaev described a simple exactly solvable model (the toric code) for a Z2 spin liquid (*31*). Building on these developments, many concrete models were constructed and reliably shown to have spin liquid phases with a variety of emergent gauge structures, in both two (*32,33,34*) and three dimensions (*34,35*). Though the matter of principle question was answered in the affirmative, the question of which of these phases, if any, occur in realistic models of materials remained largely open, and is still not satisfactorily settled.

Anderson's idea in 1973 that the near-neighbor Heisenberg model on a triangular lattice was a spin liquid turned out to be incorrect, even for spin ½ systems where quantum effects are maximized, as shown by Huse and Elser (*36*) among others, though modifications of this ideal model, for instance the inclusion of ring exchange, can still lead to spin liquid states (as we allude to below when talking about real materials such as the 2D organic ET and dmit salts). This led to the study of other lattices where antiferromagnetic interactions would be more frustrated (i.e., act to suppress long-range magnetic order). The classic example in 2D is a lattice of corner sharing triangles known as the kagome lattice (Figure 2a). In the case of a near neighbor Heisenberg model on a kagome lattice, continuous rotations of certain spin clusters are possible at no energy cost (*37,38,39*), implying a large manifold of soft fluctuation modes that act to suppress classical order.  This is dramatically evident in exact diagonalization studies (*40*), which show a spectrum of states qualitatively different from the triangular lattice case, with a dense set of both singlet and triplet excitations extending down to low energies.  Such studies have been unable to definitively address whether the excitation spectrum for both singlets and triplets are gapped or not due to limitations of modern supercomputers (the largest lattice studied so far has been 48 sites (*41*)).  Researchers have addressed larger lattices by using approximate techniques based on quantum information-like methods, such as DMRG (density matrix renormalization group) and various generalizations such as PEPS (projected entangled pair states) and MERA (multiscale entanglement renormalization ansatz).  The basic conclusion of such studies of the kagome lattice is that there are a number of states which have almost equal energies (*9*), including gapped Z2 spin liquids, gapless spin liquids (so-called U(1) spin liquids where the spinons have a Dirac-like dispersion), and long-period valence bond solids (the spin liquid ground state implied by DMRG studies (*42*) appears to be a "melted" version of a twelve-site valence bond solid that has a diamond-like structure, as shown in Figure 2b).  Although some of the latest studies point to the U(1) gapless spin liquid (*43*), it has also been shown that the latter can be unstable to anisotropy (*44*) (i.e., a striped spin liquid).  Unfortunately, exact diagonalization studies, given their limited lattice size, have not been able to resolve this matter, and indicate that the ground state might actually break inversion symmetry, or even be chiral in nature (*41*).  Also, the kagome lattice differs from the triangular lattice by the lack of a point of inversion between nearest neighbors. This admits Dzyaloshinskii-Moriya (DM) interactions between in-plane spin components that can qualitatively change the ground state relative to that of the Heisenberg model. There are indications from simulations that the addition of DM interactions indeed favors magnetic order (*45,46,47*).

In 2006, another exactly solvable model was reported by Kitaev (*48*). The lattice in this case is a honeycomb one, and the Hamiltonian is a subset of the Heisenberg model where exchange on the "x" bonds of the honeycomb only involve $S_x S_x$, on the "y" bonds only $S_y S_y$ and on the "z" bonds only $S_z S_z$ (Figure 2c).  Its ground state is a Z2 spin liquid with a gapless spectrum of fermionic ε particles.  Making the model anisotropic between the x, y, and z bonds preserves the exact solubility, but gaps out the ε particle.  The exact solution in fact yields not just the ground state, but the full spectrum of excitations. The manifold of states can be factored into flux sectors, with the flux referring to the product of the sign of the singlets around a hexagonal loop in the honeycomb (for the ground state, +1 for all hexagons).  Flux excitations are precisely the visons mentioned above, and are localized with a small energy gap.  But the "unbound" Majorana (the ε particle) is free to propagate and forms a dispersion that can be either gapped or gapless depending on the ratio of the various $J$ ($J_x, J_y, J_z$).  The visons act as a low energy momentum sink for the Majoranas, leading to a rather featureless spin excitation spectrum, as could be measured by neutrons (*49*).  One consequence is emergent fermionic statistics in the spin continuum as would be measured by Raman scattering (*50*).  Even more dramatic is the prediction of Majorana edge currents in a magnetic field, that would lead to quantization of the thermal Hall effect with a value half that expected for fermionic edge modes (*51*).

Despite the seemingly contrived form of this model, it was pointed out by Jackeli and Khaliullin in 2009 (*52*) that the model might be physically realized in certain honeycomb (and "hyper-honeycomb") iridates and related materials such as α-RuCl$_3$, which has led to an explosion of interest (Figure 2d). This brings us to our next question.

*Do quantum spin liquids really exist in Nature?* Although a spin-½ antiferromagnetic chain is a quantum spin liquid (and has long been observed in experiments), it is qualitatively different from those in higher dimensions (for instance, there is no braiding in 1D). Beyond one dimension, a number of interesting candidate materials have emerged that might host quantum spin liquids, but the evidence is circumstantial. The focus has been on materials with spins on lattices that frustrate conventional Néel order. Spin ½ systems are of particular interest since they are the least classical, but the possibility of long-range entanglement for higher spin states should not be overlooked. Fluctuations are enhanced in 2D and for low coordination numbers, but even in 3D, there are pyrochlore and hyper-kagome lattice systems that fail to develop magnetic order due to geometrical frustration. Our theoretical understanding further suggests that 'weak' Mott insulators that are close to the metal-insulator transition are fertile grounds for quantum spin liquid phases. Three of the most actively discussed classes of materials at the present time are shown in Figure 3, and all involve lattices where either the spin, s, or the total angular momentum, j, have a value of ½. They are (1) 2D organic salts such as κ-(ET)$_2$Cu$_2$(CN)$_3$ and EtMe$_3$Sb[Pd(dmit)$_2$]$_2$ where structural dimers possessing a single spin ½ degree of freedom form a triangular lattice, (2) herbertsmithite (and the closely related Zn-barlowite) where the Cu$^{2+}$ ions form a kagome lattice, and (3) α-RuCl$_3$ where the Ru$^{3+}$ ions form a honeycomb lattice. The last two are deep in the Mott insulating phase, while the organic salts are weak Mott insulators close to the metal-insulator transition. We discuss each in turn, starting with the organics.

*2D organic salts* (*53*). Although most of these salts, where structural dimers form a (distorted) triangular lattice, are magnetic at ambient pressure, there are a few that do not order. Prominent examples are κ-(BEDT-TTF)$_2$Cu$_2$(CN)$_3$ (referred to here as κ-ET), κ-(BEDT-TTF)$_2$Ag$_2$(CN)$_3$, EtMe$_3$Sb[Pd(dmit)$_2$]$_2$ (referred to here as Pd-dmit), κ-H$_3$(Cat-EDT-TTF)$_2$, and κ-(BEDT-TTF)$_2$Hg(SCN)$_2$Br. Under pressure, the first becomes superconducting, which was why it was first synthesized and studied (*54*). NMR studies show a lack of spin ordering down to temperatures well below the Curie-Weiss temperature inferred from high temperature spin susceptibility measurements. At low temperatures, the spin susceptibility χ is a constant and the heat capacity C = γT has a linear temperature dependence (*55*). Interestingly, the Wilson ratio χ/γ is within 20% of the free Fermi gas value. This suggests there are gapless spin carrying excitations despite the lack of magnetic long-range order.

In Pd-dmit, despite its insulating nature, the thermal conductivity has a metallic form at low T (κ∝T) and is magnetic field dependent (*56*). If correct, this suggests the gapless spin-carrying excitations are also mobile in this material. However, very recently this result has been re-examined in a number of dmit samples by two groups and no such metallic thermal conductivity is found (*57,58*). Moreover, in κ-ET, there is at very low T a dip in the thermal conductivity that, if taken at face value, might indicate a very small energy gap (*59*). Clearly, this emphasizes the need for striking results on spin liquid candidates to be further investigated. In κ-(BEDT-TTF)$_2$Hg(SCN)$_2$Br, heat capacity and Raman scattering indicate magnetic *and* electric dipole degrees of freedom that remain fluctuating to the lowest measured temperatures (*60*). Theoretically, the details of exactly which spin liquid is realized in these materials is not established. The experiments suggest there may be a Fermi surface of emergent fermionic spinons (at least at very low temperatures). Establishing the presence of such a neutral Fermi surface in experiments would be a great boost to our understanding. In that context, these materials (under pressure) exhibit quantum oscillations associated with their metallic Fermi surfaces. Such oscillations have not been seen in the insulating spin liquid regime, though they were looked for (*61*) given the prediction that spin liquids with a spinon Fermi surface might host such due to weak coupling of the neutral spinons to charge fluctuations (*62*). Alternate interpretations that invoke disorder to produce a heterogeneous gapless state (*63,64,65*) also deserve further experimental and theoretical exploration.

*Herbertsmithite* (*66*). The Nocera group has used mineralogy to inspire the search for spin liquids, making one wonder whether spin liquids are hiding in some long forgotten desk drawer in a museum (as in the case of the first known naturally forming quasicrystal (*67*)). Their original studies were on iron jarosites (and vanadium and chromium variants) where the magnetic ions form a perfect kagome lattice and where interesting behavior has been observed such as spin chirality (*68*). Unfortunately, these materials have long range magnetic order, and the magnetic ions are not spin ½. However, owing to larger crystal fields and spin-orbit coupling, $Ru^{3+}$ and $Os^{3+}$ jarosites are potential candidates for a j=½ kagome lattice, though synthesizing these minerals with 4d and 5d transition metal ions presents an appreciable challenge (nonetheless, they are intriguing future targets for study).

Spin-½ $Cu^{2+}$ will not go into the jarosite except in diluted form. It was at this point that the Nocera group realized that a related mineral class was known with $Cu^{2+}$, and that was herbertsmithite, $ZnCu_3(OH)_6Cl_2$, a rare mineral first identified from a mine in Chile (*69*). They were then able to synthesize this material using a hydrothermal method (*70*), and found no evidence of long range order. Since then, single crystals have been grown using a refinement of the hydrothermal technique (*71*). This has allowed for single crystal neutron scattering studies that have revealed a broad continuum of spin excitations (*72*) (Figure 4a). Surprisingly, these excitations can be described by a dynamic correlation function of the "local" form $S(\mathbf{q},\omega) = f(\mathbf{q}) g(\omega)$, reminiscent of the marginal Fermi liquid conjecture of Varma and colleagues (*73*). Such a form is not predicted by any known spin liquid model, though some resemblance can be had by using low energy visons as a momentum sink for the spinons (*74*). This brings up the important question of disorder. In particular, though it is claimed that the kagome spins are gapped (as inferred from NMR (*75*) and later neutron studies (*76*)), in reality, the entire low energy spectrum is dominated by impurity spins (often referred to as "orphan" spins). These spins are due to the fact that the transition metal sites between the kagome planes are not completely inhabited by non-magnetic Zn but also include magnetic Cu ions (*77*). Similar issues exist when Zn is replaced by other 2+ ions such as Mg or Cd. Getting rid of these impurity spins is a major challenge, not only for herbertsmithite, but for most spin liquid candidates where similar effects are also seen. This is important because some of the properties seen in herbertsmithite are reminiscent of random singlet states where one has a distribution of *J* (*78*), and it has been claimed that the inelastic neutron scattering (INS) data can be understood in this way as well (*79*). Such random states are not quantum spin liquids (since their wavefunctions have a product form) even though they do exhibit quantum-critical-like scaling.

These issues have led to the study of related materials such as Zn-barlowite, which is similar to herbertsmithite except the kagome layers are stacked differently (*80*). One advantage of Zn-barlowite is that the fluorine nuclear magnetic resonance (NMR) line is simple given its nuclear spin of ½. Analysis of these NMR data indicate a spin gap whose field dependence is consistent with a gas of spin ½ particles (i.e., spinons) (*81*) (Figure 4b). This is further supported by INS studies, which indicate the INS spin gap is twice that inferred by NMR (consistent with the fact that INS measures spin 1 excitations, i.e. pairs of spinons) (*82*), though it should be remarked that, as in herbertsmithite (*75,76*), the low energy properties of barlowite are dominated by defect spins. Most recently, attempts have been made to dope herbertsmithite to realize the long sought "doped spin liquid" popularized by Anderson in 1987. Unfortunately, intercalating Li (*83*) or replacing $Zn^{2+}$ by $Ga^{3+}$ (*84*) leads to localized polarons (as confirmed by density functional calculations (*85*)), and thus no mobile carriers. We also note a word of caution with materials produced by such reduction approaches. In using strong reducing reagents on herbertsmithite, such as alkali metals, one needs to avoid the reduction of the protons to make hydrogen instead of reducing $Cu^{2+}$. Even small amounts of deprotonated hydroxides in herbertsmithite will result in structural distortions, therefore potentially breaking crucial symmetry that might be required for stabilizing quantum spin liquid behavior. Finally, even if polarons were not to occur, DMRG simulations predict Wigner crystallization (*86*).

*α-RuCl₃* (*87*). The proposal by Jackeli and Khaliullin (*52*) that certain Mott-Hubbard systems with partially filled $t_{2g}$-levels and strong spin-orbit coupling might realize the Kitaev model led to an intense search. First studied were materials like $\alpha$-$Na_2IrO_3$ and $\alpha$-$Li_2IrO_3$ where $Ir^{4+}$ ions (with effective j = ½) form a honeycomb lattice. While these exhibit long-range order, polarized resonant x-ray data show bond-directional Kitaev interactions indeed occur in this class of materials (*88*). This is why the recent discovery of a variant, $H_3LiIr_2O_6$, that does not exhibit long range order, is important (*89*).

The realization that α-RuCl₃ has similar properties to the iridate honeycomb materials led to a huge growth in these studies. In α-RuCl₃, magnetic Ru is found on a honeycomb lattice between close packed Cl planes. This material is not only relatively easy to grow in single crystalline form and manipulate (as the layers are van der Waals coupled, they can be exfoliated), the thermal neutron absorption cross section for Ru is 170 times less than for Ir, so α-RuCl₃ is amenable to INS studies, which reveal a high energy continuum of spin excitations (*90*). On the other hand, there has been some debate about which properties of this material are attributable to Kitaev physics, as opposed to more traditional physics (these materials have a non-negligible Heisenberg interaction). In particular, whether magnon-like excitations could explain some (or all) of the data (*91*), realizing that the material does order at low temperatures. Still, the spin continuum as detected in Raman data seems to obey fermionic statistics (*50*). Most dramatically, realizing that magnetic order is suppressed upon applying a magnetic field, Matsuda's group was able to detect a thermal Hall signal that plateaued in a small range of temperature and field (*92*) (Figure 4c). The value of this plateau is consistent with Majorana edge modes, being one-half of the value for fermionic edge modes (*51*). Seeing such a plateau is peculiar given that the thermal Hall angle is small (the longitudinal thermal conductivity is dominated by phonons), but this has been explained by two different theory efforts (*93,94*). As with most important experiments in this field, this result has yet to be reproduced by other groups. And, consistent with the organics and herbertsmithite, disorder should play a significant role as well, particularly due to the presence of stacking faults (*95*). Finally, based on the evidence that α-RuCl$_3$ exhibits spin liquid behavior, it is of great interest to study the physical properties of electron/hole doped variants (*96*). It is worth noting that while doping α-RuCl$_3$ via chemical reduction or oxidation, one should always be careful to perform the reaction under an inert and non-nitrogen atmosphere because Ru complexes react with nitrogen.

A big question about Kitaev materials is "Is the Kitaev model relevant to Kitaev materials"? The spin liquid in the exact solution may have only a tiny regime of stability beyond the solvable limit, based on numerical calculations of the Kitaev model supplemented with Heisenberg exchange interactions (*97*). Furthermore, in the exact solution, the vison gap is very small (only a few percent of the Kitaev exchange) and so thermally the spin liquid state only occurs at very low T. Recent calculations suggest that a certain spin-anisotropic "symmetric exchange" enhances the stability of the exactly solved spin liquid (*98*). Alternately, the possibility that any spin liquid that occurs in α-RuCl$_3$ or the iridates may not be smoothly connected to the Kitaev spin liquid must be kept in mind (*99*).

*An embarrassment of riches.* Space precludes us from saying something more than cursory about other spin liquid candidates, noting that nowadays there are many claims of such. Of recent interest has been YbMgGaO$_4$ where the Yb ions form a triangular lattice, albeit with disorder on the non-magnetic cation site. It is easy to grow and study, and the fact that the energy scales associated with the 4f Yb ion are small makes it more amenable to certain types of studies (extensive neutron scattering studies have been done (*100*)). It too has been claimed to possibly have a "spinon" Fermi surface (*101*), but as with most spin liquid candidates, disorder plays a significant role (*64,102*) (due in this case to Mg and Ga interchanges that distort the Yb environment (*103*)). Another candidate, Ca$_{10}$Cr$_7$O$_{28}$, can be described as a triangular lattice of six Cr$^{5+}$-based spin-1/2 clusters – each consisting of an antiferromagnetic and a ferromagnetic triangle interacting ferromagnetically with each other. Extensive experimental and numerical work on this bilayer kagome material has established its spin Hamiltonian and the lack of static spin ordering down to 0.3 K (*104*). For both YbMaGaO$_4$ (*105*) and Ca$_{10}$Cr$_7$O$_{28}$ (*106*), however, the absence of a linear term in the thermal conductivity argues against the existence of a spinon Fermi surface. Recently, a copper oxide, averievite, Cu$_5$V$_2$O$_{10}$(CsCl), has been found where the copper ions form a pyrochlore slab. First discovered in a volcano in Kamchatka, the material was synthesized and subsequently buried in a thesis, only to be "rediscovered" (thanks to Google Scholar) (*107*). Substitution by zinc likely replaces the inter-site copper ions (as in herbertsmithite), isolating the copper kagome layers, and the resulting susceptibility and specific heat are reminiscent of herbertsmithite (*107*). Several materials are also known where magnetic ions form a "hyper-kagome" lattice (obtained by taking the kagome layer and pulling it into the third dimension). Of particular note are Na$_4$Ir$_3$O$_8$ (*108*) and PbCuTe$_2$O$_6$ (*109*), but again both have quenched disorder (in the former case due to partial occupation of the Na sites) and distortions (in the latter case, there are many exchange parameters associated with its distorted hyper-kagome lattice), which make these materials less than ideal. As for other frustrated 3D lattices, extensive studies on rare earth and transition metal pyrochlores would take us well beyond the scope of this article, and the reader is referred to a recent review (*110*). An exciting recent proposal (*111*) which seems to be receiving some experimental support (*112*) is that the layered transition metal dichalcogenide 1T-TaS$_2$ might be a quantum spin liquid.

*And potential riches*. The above may leave one to ask "What else is out there?" Almost certainly a lot. Many interesting materials known in mineralogical form have yet to be made in the lab and studied for their magnetic properties. As an example, quetzalcoatlite (named after an Aztec god) has coppers on a perfect kagome lattice, and was the first mineral whose crystal structure was determined at the Advanced Photon Source (*113*). But it, like many other minerals, is only known by its structure and nothing else. Certainly, a systematic study of all promising minerals (i.e., those whose magnetic ions sit on a frustrated lattice) would be a good start. If something interesting is found, then an attempt to synthesize it would be in order. A recurrent challenge with frustrated magnets is that chemical disorder acts at the "ultraviolet" scale, giving rise to orphan spins. Clearly more attention (and resources) needs to be devoted to synthesis, both in developing promising new synthesis routes (high pressure, hydrothermal, molecular beam expitaxy, etc.), and finding ways to mitigate and control disorder. This is a tall order (single crystals of herbertsmithite, for instance, take months to grow, and after all that, are still small and non-stoichiometric). Substitutional studies are also warranted. We mentioned above the tantalizing possibility of replacing Fe by Ru or Os in jarosites. Similarly, one wonders what the osmium analogue of $\alpha$-RuCl$_3$ would be like. And, of course, obtaining a doped spin liquid that is metallic would be the holy grail for many (*14,114*). This could potentially be accomplished by ionic liquid gating to avoid chemical disorder.

*The Future*. As the old saying goes, "It's tough to make predictions, especially about the future". Nonetheless, several observations are in order. Having addressed materials-based issues above, we turn to theory. Although great strides have been made in numerical techniques, we still do not know, for instance, what the ground state is of the near-neighbor Heisenberg model on a kagome lattice, and less about many other frustrated lattices, or for "real" Hamiltonians that contain multiple exchange parameters as well as anisotropic exchange and DM terms. Still less is known about dynamical and non-equilibrium properties. While neutron scattering when combined with theoretical calculations of S($\mathbf{q}$,$\omega$) can provide circumstantial evidence for a spin liquid, methods to probe entanglement are needed to establish model-independent evidence. As spin liquids are spin relatives of the fractional quantum Hall effect, applying methods known from spintronics to study them would be in order. In particular, to search for spin currents (*115,116*), the spin Hall effect, spin noise, and other spin-related effects that might expose the nature of the spinons (if they indeed exist). As for visons, a proposal for their study was made many years ago by one of us (*117*) that involves looking for trapped flux in a spin-liquid ring. This experiment was actually performed on a superconducting cuprate with a null result (*118*), but obviously doing this sort of experiment on spin liquid candidate materials would be in order. Similarly, impurities can be exploited not only to trap Majorana fermions, but also Friedel oscillations about defects (that could be detected by a scanning tunneling probe) could reveal a spinon Fermi surface if such should exist (*119*). And tunneling has been advocated as a possible probe of how electrons could potentially fractionalize when injected into a spin liquid (*120*). Ultimately, if topological excitations were identified in a material, then one could think about probing and extending their phase coherence time and braiding them in steps towards their utilization for "topological" quantum computation (*121*) (Figure 1c). As for other potential applications, we can think of no better way to end than with Michael Faraday's response to William Gladstone's dismissal of a scientific discovery: "What use is it?", he quipped. "Why, sir, there is every probability that you will soon be able to tax it".

**Acknowledgments.** The authors would like to thank Maria Hermanns, Patrick Lee, Young Lee, and Shivaji Sondhi for their comments.

**Funding.** C.B. and R.J.C. were supported by the U.S. Department of Energy, Office of Science, Basic Energy Sciences, through DE-SC-0019331. S.A.K. was supported by the National Science Foundation grant DMR-1608055. M.R.N. was supported by the U.S. Department of Energy, Office of Science, Basic Energy Sciences, Materials Sciences and Engineering Division. T.S. was supported by the National Science Foundation grant DMR-1608505, and partially through a Simons Investigator Award.

**Author Contributions.** All authors contributed to the text.

**Competing Interests.** The authors declare no competing interests.

**Correspondence.** Correspondence should be addressed to M.N. (norman@anl.gov).

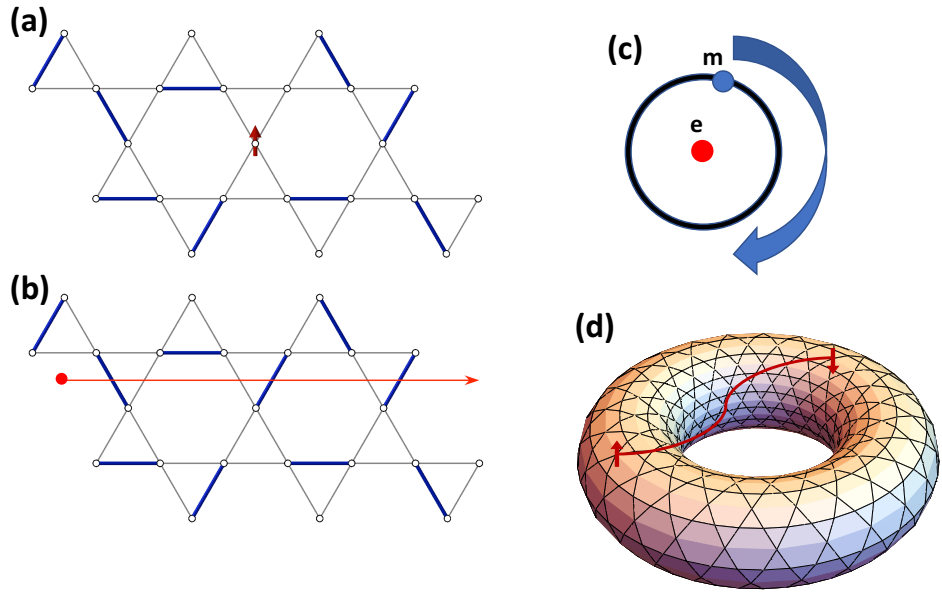

Figure 1: **Excitations of a Spin Liquid.** Cartoon of (a) a spinon excitation, (b) a vison excitation, and (c) a braiding of anyons. Blue bonds represent singlets, the red arrow in (a) a spinon, the red line with an arrow in (b) a vison (the sign of the singlet changes for every one intersected by this line), and e and m in (c) denote anyons. (d) illustrates long-range entanglement of two spins, with the torus representing the ground state degeneracy typical for gapped spin liquids (the Z2 spin liquid has a degeneracy of four on the torus associated with the topologically distinct horizontal and vertical loops that encircle the torus).

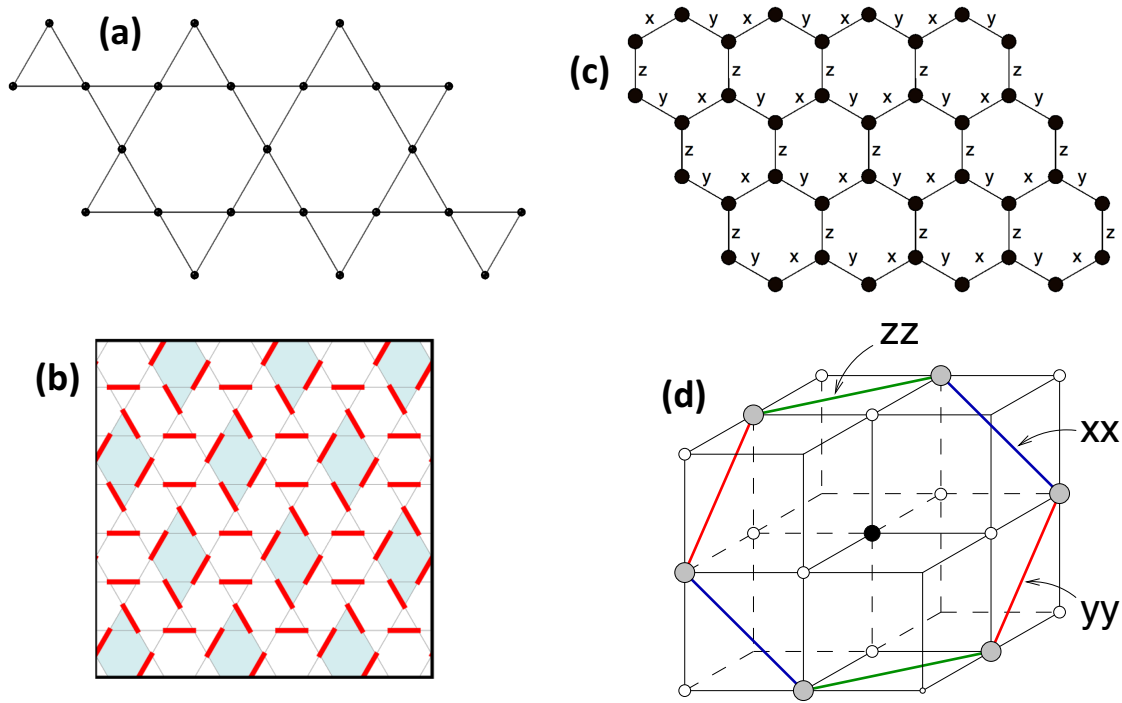

Figure 2: **Geometrically Frustrated Models.** (a) kagome lattice, (b) diamond valence bond solid on a kagome lattice (*122*), (c) Kitaev model on a honeycomb lattice, and (d) bond-dependent Kitaev interaction in a six-fold coordinated transition metal oxide (*52*). In (b), red bonds are singlets, with blue shading emphasizing the diamonds. In (c) and (d), x (xx), y (yy), z (zz) denote the component of the spins involved in that bond.

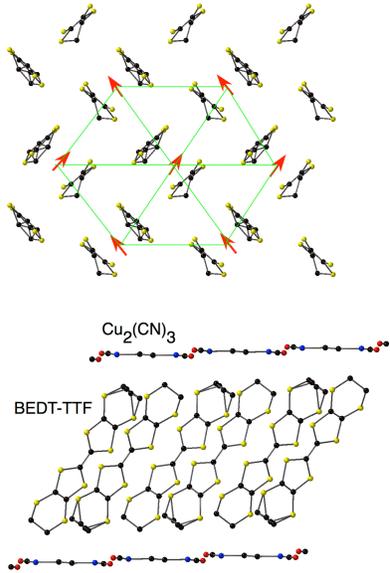 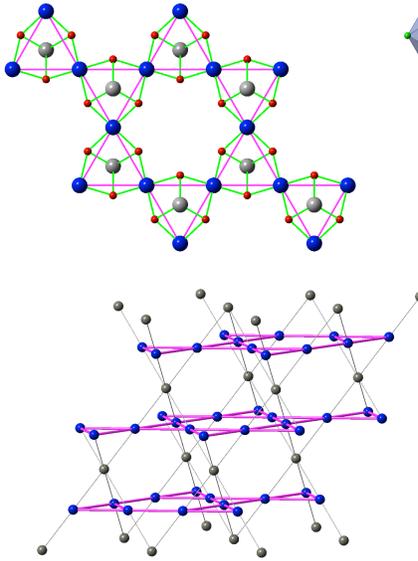 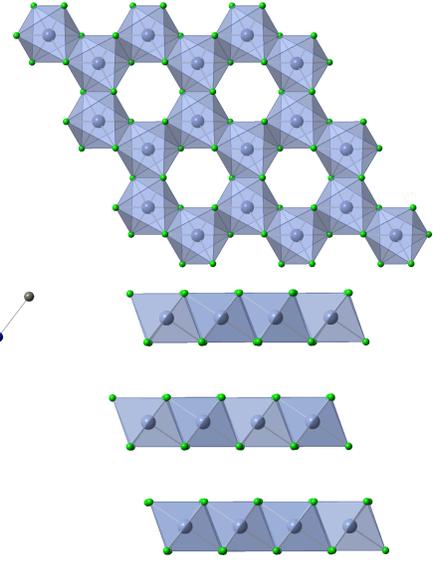

Figure 3: **Candidate Spin Liquid Materials.** Crystal structures of (a) $\kappa\text{-}(ET)_2Cu_2(CN)_3$, (b) herbertsmithite, and (c) $\alpha\text{-}RuCl_3$. In (a), the ET dimers (top) form a triangular lattice (with the S = ½ degree of freedom per dimer represented by red arrows). These ET molecules are sandwiched by $Cu_2(CN)_3$ planes (bottom). Here, Cu is in blue, S in yellow, C in black, and N in red. In (b), Cu (blue) forms kagome layers (top) that are interconnected (bottom) by Zn (gray), with O in red (shown in top only). In (c), Ru octahedra (top) form honeycomb layers that are weakly coupled (bottom), with Cl in green.

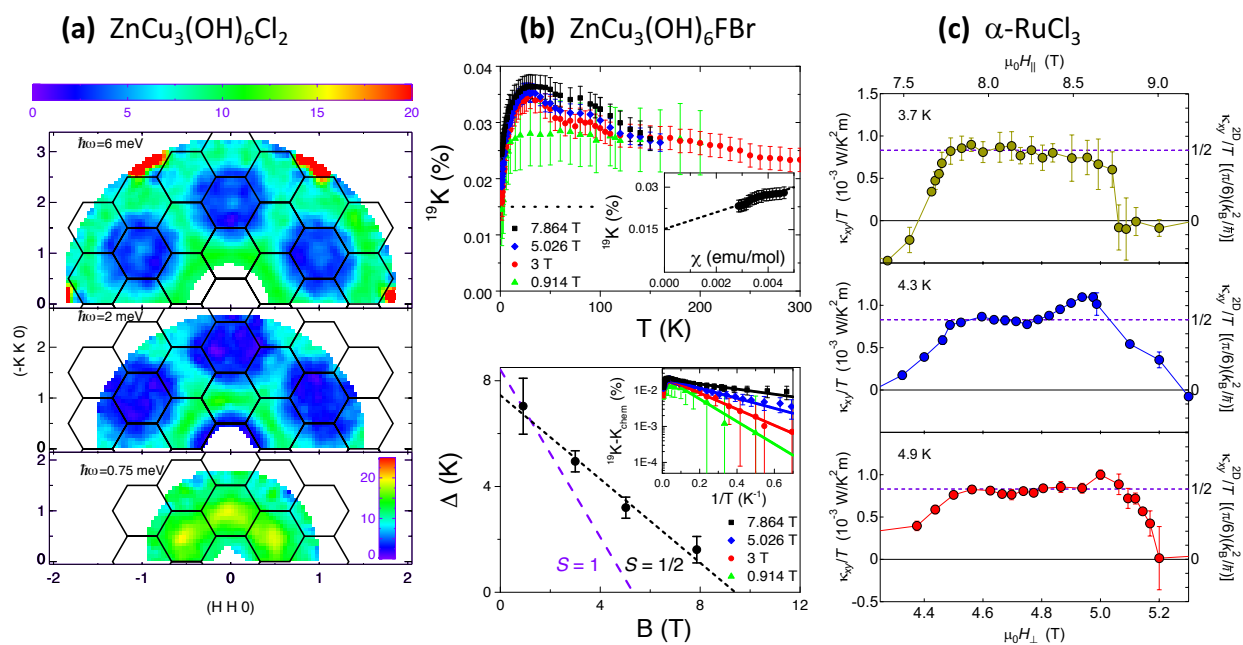

Figure 4: **Key Data on Spin Liquid Candidates.** (a) Spin continuum of herbertsmithite from inelastic neutron scattering [$S(\mathbf{q},\omega)$ at 1.6 K in the HK0 plane: upper, 6 meV; middle, 2 meV; lower, 0.75 meV] (*72*), (b) field dependence of the spin gap of Zn-barlowite from nuclear magnetic resonance [upper: $^{19}$F Knight shift versus temperature for various magnetic fields; lower: magnetic field dependence of the spin gap, $\Delta$, with dashed lines the expected behavior for $S=½$ and $S=1$ excitations] (*81*), and (c) quantized plateau in the thermal Hall effect of $\alpha$-RuCl$_3$ [$\kappa_{xy}/T$ versus magnetic field: upper, 3.7 K; middle 4.3 K; lower, 4.9 K] (*92*).